\documentclass[10pt,journal,compsoc]{IEEEtran}

\usepackage{gensymb}
\usepackage{cite}
\usepackage{graphicx}
\usepackage{amsmath}
\usepackage{amssymb}
\usepackage{amsthm}
\usepackage{epstopdf}
\usepackage{color}
\usepackage{latexsym,amssymb,epsfig,color,verbatim}
\usepackage{hyperref}
\usepackage{subcaption}
\usepackage{multirow}
\usepackage{wasysym}
\usepackage{xcolor}
\usepackage{tabu}
\usepackage[ruled,vlined,linesnumbered]{ algorithm2e}
\usepackage[normalem]{ulem}
\usepackage{pbox}

\usepackage{tikz}
\usetikzlibrary{arrows.meta,shapes,graphs,graphs.standard,quotes}
\tikzset{>=latex}
\tikzstyle{vertex}=[circle,fill=black!25,minimum size=10pt,inner sep=0pt]
\tikzstyle{arc} = [draw,thick]

\newcommand{\blue}[1]{\textcolor{blue}{#1}}

\newtheorem{definition}{Definition}

\begin{document}

\title{\blue{Learning Features from Graphlets}}
\title{Learning Features of Network Structures \\Using Graphlets}

\author{Kun~Tu$^*$, %~\IEEEmembership{Member,~IEEE,}
        Jian~Li$^*$,~\IEEEmembership{Member,~IEEE,}
        Don~Towsley,~\IEEEmembership{Life~Fellow,~IEEE,}
       Dave~Braines,~\IEEEmembership{Member,~IEEE,}
        and~Liam~D.~Turner,~\IEEEmembership{Member,~IEEE}% <-this % stops a space
        \thanks{$^*$Co-primary authors.}
        \thanks{This research was sponsored by the U.S. Army Research Laboratory and the U.K. Ministry of Defence under Agreement Number W911NF-16-3-0001. The views and conclusions contained in this document are those of the authors and should not be interpreted as representing the official policies, either expressed or implied, of the U.S. Army Research Laboratory, the U.S. Government, the U.K. Ministry of Defence or the U.K. Government. The U.S. and U.K. Governments are authorized to reproduce and distribute reprints for Government purposes notwithstanding any copyright notation hereon.  }%The material in this paper was presented in part at the IEEE/ACM International Conference on Advances in Social Networks Analysis and Mining (ASONAM), Vancouver, Canada in 2019 \cite{kun19asonam}.}
\IEEEcompsocitemizethanks{\IEEEcompsocthanksitem K. Tu is with the College of Information and Computer Sciences, University of Massachusetts Amherst, Amherst, MA, USA. %\protect\\
E-mail: kuntu@cs.umass.edu.
\IEEEcompsocthanksitem J. Li is with the Department of Electrical and Computer Engineering, Binghamton University, the State University of New York, Binghamton, NY, USA. %\protect\\
E-mail: lij@binghamton.edu.
\IEEEcompsocthanksitem D. Towsley is with the College of Information and Computer Sciences, University of Massachusetts Amherst, Amherst, MA, USA. %\protect\\
E-mail: towsley@cs.umass.edu.
\IEEEcompsocthanksitem D. Braines is with IBM Research UK, Emerging Technology, Hursley Park, Winchester, UK. 
E-mail: dave${}\_{}$braines@uk.ibm.com.
\IEEEcompsocthanksitem L. D. Turner is with School of Computer Science and Informatics, Cardiff University, Cardiff, UK. E-mail: TurnerL9@cardiff.ac.uk.}% <-this % stops an unwanted space
}

\IEEEtitleabstractindextext{

\begin{abstract}
Networks are fundamental to the study of complex systems, ranging from social contacts, message transactions, to biological regulations and economical networks.  In many realistic applications, these networks may vary over time.  Modeling and analyzing such temporal properties is of additional interest as it can provide a richer characterization of relations between nodes in networks.  In this paper, we explore the role of \emph{graphlets} in network classification for both static and temporal networks.  Graphlets are small non-isomorphic induced subgraphs representing connected patterns in a network and their frequency can be used to assess network structures.  We show that graphlet features, which are not captured by state-of-the-art methods, play a significant role in enhancing the performance of network classification.  To that end, we propose two novel graphlet-based techniques, \emph{gl2vec} for network embedding, and \emph{gl-DCNN} for diffusion-convolutional neural networks.  We demonstrate the efficacy and usability of \emph{gl2vec} and \emph{gl-DCNN} through extensive experiments using several real-world static and temporal networks.  We find that features learned from graphlets can bring notable performance increases to state-of-the-art methods in network analysis.
\end{abstract}

}

\maketitle

\IEEEdisplaynontitleabstractindextext

\IEEEpeerreviewmaketitle

\section{Introduction}\label{sec:intro}

Networks, where elements are denoted as nodes and their interactions are denoted as edges, are fundamental to the study of complex systems \cite{albert02,newman10}, including social, communication, and biological networks.  Analysis of networks include network classification, community detection and so on.  Typical analysis usually models these systems as static undirected graphs that describe relations between nodes.  However, in many realistic applications, these relations are directional and may change over time \cite{holme12,kovanen11,paranjape17}.  Modeling these \emph{directed and temporal} properties is of additional interest as it can provide a richer characterization of relations between nodes in networks.  It has become one of the most in-demand but computationally challenging problems in today's highly connected society.

\begin{figure}
\centering
    \includegraphics[width=0.8\linewidth]{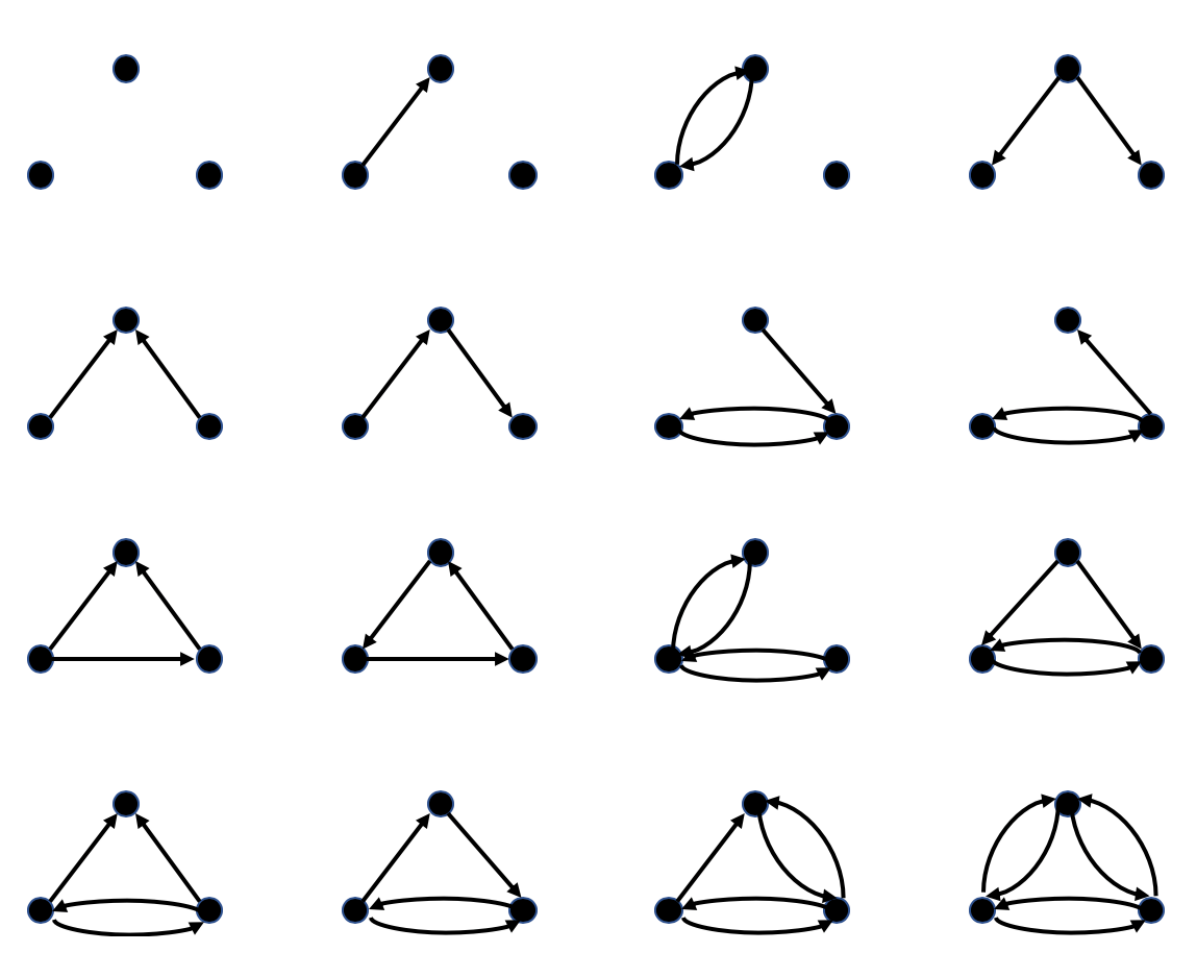}
  \caption{All 16 triads in static directed network \cite{doroud2011evolution}.}
  \label{fig:static-motifs}
\end{figure}

\begin{figure}
\centering
 \includegraphics[width=0.8\linewidth]{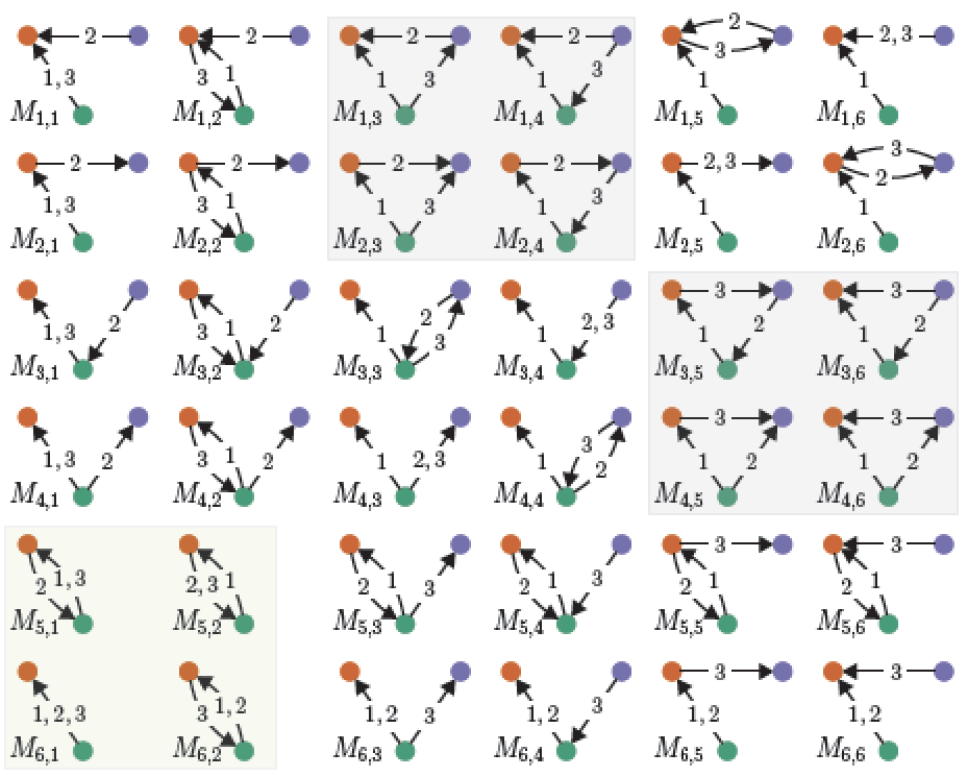}
  \caption{All $2$-node and $3$-node, $3$-edge, $\delta$-temporal graphlets as defined in \cite{paranjape17}. Edge labels correspond to the ordering of edges. All $36$ graphlets are labeled with $M_{i,j}$ across $6$ rows and $6$ columns. The first edge in each graphlet is from the green to the orange node. The second edge is the same along each row, and the third edge is the same along each column.}
  \label{fig:temporal-motifs}
\end{figure}

A wide variety of analytical approaches have been proposed for network classification tasks in networks.  This often involves applying machine learning techniques to these problems, which requires the network to be represented as a feature vector.  However, representing a network as a feature vector is challenging due to high dimensionality and potentially large and complex network structures.  In this paper, we propose to advance the state-of-the-art by exploring the role of \emph{graphlets} in network classification for both static and temporal networks.  Graphlets are small non-isomorphic induced subgraphs representing connected patterns in a network and their frequency can be used to assess network structure. For example, Figure~\ref{fig:static-motifs} shows all $16$ triads in static directed network, and Figure~\ref{fig:temporal-motifs} shows all possible $2$-node and $3$-node, $3$-edge, $\delta$-temporal graphlets.  These will be described in detail in Section~\ref{sec:formulation}.

While there is a dramatic increase of research into networks in recent years, there is little work investigating graphlets within graphs that capture important human ties and interactions among these networks.  Additionally, most studies of graphlets focused on their presence in \emph{static} graphs and little is known about how they change over time.  Such an understanding of network dynamicity can offer interpretable explanations into the possible role or purpose of the observed network based on similarity to other networks.  Such analyses can also provide guidance on addressing temporal changes in networks on network analysis, including comparison, classification and prediction of temporal network behaviors.   In this paper, we focus on investigating the functionality of graphlets in network embedding and graph neural networks for network classification.

\noindent\textbf{Network Embedding.}
Various ways of learning \emph{feature representations} of nodes in networks have been recently proposed to exploit their relations to vector representations \cite{shervashidze2011weisfeiler, grover2016node2vec, ribeiro17,adhikari2018sub2vec, narayanan2017graph2vec}.  However, most of these are applied to node and edge predictions and fail to fully capture \emph{network structures}.  It remains an open question as to whether network classification through node embedding methods can be improved, since the whole network structure clearly also plays a significant role.   

\noindent\textbf{Graph Neural Networks.}  Graph neural networks (e.g., graph convolutional networks \cite{kipf2016semi}, diffusion convolutional neural networks (DCNNs)  \cite{atwood2016diffusion}) has been proposed as a new model to learn representations from graph-structured data, which can outperform probabilistic relational models and kernel methods at node classification tasks.  Graph neural networks, in particular, DCNNs,  offer a complementary approach that provides a significant improvement in predictive performance at node classification tasks.  However, its benefit on network classification tasks is not clear.

In this paper, we demonstrate the benefit of including graphlet features in network embedding and graph neural networks for network classification.  To achieve this goal, we proposal a graphlet-based network embedding methodology, {\emph{gl2vec}}, and a graphlet-based DCNN, {\emph{gl-DCNN}}, respectively, and apply these in extensive experiments to show significant performance improvements.

\subsection{gl2vec}
We propose  a novel network embedding methodology, {\emph{gl2vec}}, to address the aforementioned issues with current state-of-the-art techniques.  To capture network structure, \emph{gl2vec} constructs vectors for \emph{feature representations} by comparing \emph{static or temporal network {graphlet} statistics in a network to random graphs generated from different null models} (known as subgraph ratio profile, i.e., SRP, see Section~\ref{sec:embedding}).  Null models are used to generate random graphs with specific structural features to determine the significance of graphlet frequency with a specific graph (See Section~\ref{sec:embedding}).  We show how the ratios of occurrences of graphlets in a network to their occurrences in random graphs (SRPs) can be used as a fixed length feature representation to classify and compare networks of varying sizes and periods of time with improved accuracy.

We demonstrate the efficacy and usability of {\emph{gl2vec}} on network classification tasks such as network type classification and subgraph identification in several real-world static and temporal directed networks.  We apply various well-known machine learning models along with our graph feature representation for network classifications, and make a comparison with state-of-the-art methods, such as different graph kernels \cite{shervashidze2011weisfeiler}, \emph{node2vec} \cite{grover2016node2vec}, \emph{struc2vec} \cite{ribeiro17}, \emph{sub2vec} \cite{adhikari2018sub2vec}, and \emph{graph2vec} \cite{narayanan2017graph2vec}.

First, we study how static and temporal network {graphlets} can be used to classify the \emph{network type}.   A network type is defined as the network domain \cite{paranjape17}  (or superfamily \cite{milo04}, representing the type of relation or interaction between nodes) that a network belongs to, e.g.,  email networks, Google+ or Twitter in social networks, question answering networks, or even networks representing switching between mobile apps.  Graphs or subgraphs from the same network type often have similar structures \cite{milo04}.  Correctly identifying the type of network further allows us to study interactions between nodes, and predict unobserved network structures.   Secondly, we consider the problem of identifying a particular (sub)network within the same network type from its static or temporal topological structure.  For example, we predict the community ID for (sub)graphs within a network, such as identifying a department based on the temporal email-exchange pattern or detecting a mobile phone user given their app switching behaviors represented as static or temporal networks.

Given a network topological structure, identifying the network type or a network community ID in a network can be viewed as a (sub)graph classification problem. Many existing methods use different graph embedding techniques to represent graphs in a vector space and apply machine learning methods for classification. Yet, little work has applied network {graphlets} to real-world application in directed (temporal) network classification.  In this paper, we find that a strong relation exists between network type, graphlet distribution and subgraph ratio profile (SRP), a measure to decide if a {graphlet} is a significant pattern in a network by comparing {graphlet} counts between an empirical network and random graphs from a null model.  We also numerically characterize the impact of different null models on the performance of network classification in static directed networks and results show that all models achieve equally good performance.

From this, we find that {\emph{gl2vec}} identifies additional network features that are not captured by state-of-the-art methods.  These features are important to improve the accuracy of network classification in both static and temporal directed networks.  We show that concatenating {\emph{gl2vec}} to aforementioned state-of-the-art methods for network type classification and subgraph identification tasks can bring notable performance increases.

\subsection{gl-DCNN}
DCNN requires node attributes to embed nodes and graphs.  We propose, \emph{gl-DCNN}, which concatenates additional graphlet features to DCNN.  We demonstrate the efficacy and usability of {\emph{gl-DCNN}} on both node classification and graph classification in several real-world networks.  We show that \emph{gl-DCNN} can significantly improve the performance of DCNN, especially in graph classification.  

The rest of the paper is organized as follows.  Firstly, we present preliminaries in Section~\ref{sec:formulation}.  From this, we present \emph{gl2vec} and \emph{gl-DCNN}, and their evaluations in Section~\ref{sec:embedding}, and Section~\ref{sec:gcnn}, respectively. We discuss related work in Section~\ref{sec:related} and conclude our findings in Section~\ref{sec:conclusion}.

\section{Definitions}\label{sec:formulation}

In this section, we provide definitions used in the rest of the paper.  
For temporal networks and temporal network graphlets, we consider definitions given in \cite{paranjape17}, although we can equally use definitions in \cite{kovanen11}. We present them here for completeness.

\begin{definition}\label{def:temporal-network}
A \emph{temporal directed network} \cite{paranjape17} is a set of nodes and a collection of directed temporal edges with a timestamp on each edge.  Formally, a temporal directed network $T$ on a set of nodes $V$ is a collection of tuples $(u_i, v_i, t_i),$ $i=1,\cdots, N,$ where $N$ is the number of directed temporal edges, $u_i, v_i\in V$ and $t_i\in\mathbb{R}$ is a timestamp.  We refer to $(u_i, v_i, t_i)$ as a \emph{temporal edge}. 
\end{definition}
In order to strictly order the tuples, we assume timestamps $t_i$ are unique.  This assumption can be easily extended to cases where timestamps are not unique at the cost of complex notation.

\begin{definition}\label{def:static-network}
A directed static network $G(V, E)$ is defined as a set of nodes, denoted as $V$ and a set of directed edges without timestamps, denoted as $E\subset V^2\setminus\{(u,u):u\in V\}$. 
\end{definition}

In the following, we formalize the definitions of (static) graphlet and temporal graphlet.
\begin{definition}\label{def:graphlet}
Graphlets are small connected non-isomorphic induced subgraphs of a larger network.
\end{definition}

In particular, we focus on triads, shown in Figure~\ref{fig:static-motifs}.  Note that the first three triads are not connected, hence do not satisfy the graphlet definition, but we argue that they are also important in constructing vectors for network feature representation. 

It is worth mentioning that four- or even five-node graphlets can also be used for network embedding in our model.  However, counting four-or five-node graphlets are much more computationally expensive and some of our graph datasets (see Section~\ref{sec:exp}) are too large to reasonably compute counts of graphlets with more than three nodes. Given this and that triads have been widely studied in literature, we focus on triads in this work.

\begin{definition}\label{def:temporal-motif}
Temporal network {graphlets} \cite{paranjape17} are defined as induced subgraphs on sequences of temporal edges.  Formally, a $k$-node, $l$-edge, $\delta$-temporal graphlet is a sequence of $l$ edges, $M=(u_1, v_1, t_1), \cdots, (u_l, v_l, t_l)$ that are time-ordered within a  duration $\delta$, i.e., $t_1<\cdots<t_l$ and $t_l-t_1\leq\delta,$ such that the induced static graph from edges is connected with $k$ nodes. 
\end{definition}
We consider all $2$-node and $3$-node, $3$-edge, $\delta$-temporal graphlets, as shown in Figure~\ref{fig:temporal-motifs}.  Note that \cite{paranjape17} used the term network motif.

\section{gl2vec: Network Embedding with Graphlets}\label{sec:embedding}

In this section, we consider and characterize the significant importance of graphlets in designing network embedding techniques.  Network embedding has received considerable attention due to its effect on the performance of network classification, see Section \ref{sec:related}.  However, previous work has primarily focused on examining this for undirected static networks.  Applying these existing techniques to directed static networks may lose network structure information, while applying them to temporal networks may lose temporal information, and both may result in reduced accuracy.  Therefore, we introduce a new static (temporal) network embedding technique based on static (temporal) network {graphlets}.

\subsection{Subgraph Ratio Profile and Null Models}

Graph embeddings need to be independent of network size and, if temporal, the time period the network covers. While previous work has shown that the counting and probability distribution of {graphlets} are strongly related to network types \cite{paranjape17}, {graphlet} counts may differ across networks.  To achieve this independence, we use subgraph ratio profile (SRP) for network embedding, which is computed using {graphlet} counts from both the network in question and random graphs produced using a null model.
\begin{definition}
A null model \cite{newman04} is a generative model used to generate random graphs that matches a specific graph in some of its structural features such as the degrees of nodes or number of nodes and edges.
\end{definition}

\subsubsection{Static Networks}
For static networks, we consider three different null models:
 
(i) {\bf\emph{NE}:} random graphs with the same number of nodes and edges;  

(ii) {\bf\emph{MAN}:} random graphs with the same numbers of (M)utual, (A)symmetric and (N)ull edges; 

(iii) {\bf\emph{BDS}:} random graphs with the same in/out degree-pair sequence (also called bidegree sequence, BDS).   

{\bf\emph{NE}} has been widely used in previous studies since it is easy to generate random graphs \cite{Kretzschmar1996Measures} and the probability of a node degree in a random graph can be approximated by Poisson distribution in the large limit of graph size \cite{newman2001random}.  Thus network features and {graphlet} statistics can be easily modeled.  Recent studies \cite{Amaral2000Smallworld, newman2001random} showed that node degrees in a wide range of real-world networks do not necessarily follow a Poisson distribution and suggested a null model with controlled node degree sequence for network study.  We extend this to {\bf\emph{BDS}} for our study.   A study using directed networks \cite{jiang2015reciprocity} discovered that reciprocity between nodes in different social networks tends to reach maximal reciprocity constraint given in/out degree sequence.  Since reciprocities can be computed by the numbers of mutual, asymmetric and null edges, the effect of {graphlets} statistics in {\bf\emph{MAN}} on network classification is worth investigating. We numerically characterize the impact of null models on the performance of network classification in Section~\ref{sec:exp}.

\subsubsection{Temporal Networks} 
For temporal networks, since there is no equivalent null model, we consider ensembles of randomized time-shuffled data as a temporal null model \cite{mellor18}.  To be more specific, we randomly permute the timestamps on the edges while keeping the node pairs fixed.   This model breaks the temporal dependencies between edges but preserves the network structure.

In our study, we use a null model to compare {graphlet} counts in a network against random graphs. The difference between counts is then used to construct an SRP as a feature representation of the network.

\begin{definition}
Subgraph ratio profile (SRP) \cite{milo04} for a {graphlet} $i$ is defined  as 
\begin{align}
SRP_i=\frac{\Delta_i}{\sqrt{\sum \Delta_i^2}},
\label{eq:srp}
\end{align}
where 
\begin{align*}
\Delta_i=\frac{N_{ob_i}-<N_{rand_i}>}{N_{ob_i}+<N_{rand_i}>+\epsilon}.
\end{align*}
Here $N_{ob_i}$ is the count of {graphlet} $i$ observed in an empirical network, and $<N_{rand_i}>$ is the the average count in random networks in a null model.  Last, $\epsilon$ (usually set to four) is an error term to make sure that $\Delta_i$ is not too large when a {graphlet} $i$ rarely appears in both  empirical and random graphs.
\end{definition}
{A large positive value of an SRP indicates that a graphlet occurs much more frequently in a network than would be expected by random chance.}  Since SRP for a {graphlet} has been normalized, it can be used to compare different size networks. The network embedding is a vector containing $16$ SRPs for static triads. For null models of temporal directed networks, we randomly order of temporal edges. The embedding contains the SRPs for the $36$ temporal graphlets illustrated in Figure~\ref{fig:temporal-motifs}.

\subsection{Problem Formulation}
We use two network classification tasks, network type classification and subgraph identification, as the basis for evaluating \emph{gl2vec}. 

Denote $\{G_i(V_i, E_i, L_i)\}_{i=1}^N$ as (sub)graphs in different static or temporal networks, where $V_i$ is a set of nodes and $E_i$ is a set of edges in $G_i$.  If $G_i$ is a temporal network, $E_i$ is then a temporal edge with a timestamp as defined in Definition~\ref{def:temporal-network}, otherwise, $E_i$ is a directed edge.   Suppose that graphs can be categorized into $D$ classes, $D<N$.  We associate each graph $G_i$ with a label $L_i\in\{1,\cdots, D\}$.

Let $f: \{G_i\}\rightarrow \mathbb{R}^m$ be a \emph{mapping function} (also called graph embedding function) from $G_i$ to a $1\times m$ \emph{feature representation vector} defined using SRPs of static or temporal graphlets.  

Let $g: \mathbb{R}^m \rightarrow P\in \mathbb{R}^D$ be a \emph{classifier} that maps a feature representation to a categorical distribution $P$ for $D$ labels.  We represent probability distribution of $G_i$'s label as $P_i=[p_{i,1}, \dots, p_{i,D}] = g(f(G_i))$. 
 
Our goal is to solve this classification problem by designing an embedding function  $f$ and selecting a machine learning model $g$ that minimizes \emph{the sum of cross entropy \cite{de05} for all graphs}
\begin{align*}
&\arg\min_{g,f} \left(-\sum_{i}\sum_{j=1}^D\mathbf{1}_{L_i=j} \log(p_{i,j})\right) \nonumber\displaybreak[0]\\
= &\arg\min_{g,f} \left(-\sum_{i}\log(p_{i,L_i})\right).
\end{align*}
We obtain $g$ by training machine learning models.  In the next section, we discuss how to design an embedding function $f$ for static and temporal networks using graphlets.

\begin{algorithm}
	\KwData{Static or temporal graph edges list $E $,\\ \qquad \quad Null model $M$ } 
		\KwResult{Graph feature vector $\vec{f}$ }
		$\vec{N}_{ob} = $ getGraphletCounts($E$) \;
		$\vec{N}_{rand}$ = getAvgGraphletCountsInNullModel($E, M$)
		
	\For{$i = 1 : |\vec{N}_{obs}|$}{
		$\vec{f}_i$ = getSRP($\vec{N}_{obs_i}, \vec{N}_{rand_i}$)
		
	}
	\textbf{return} $\vec{f}$
	\caption{{\emph{gl2vec}}}
	\label{alg:bdsAlg}
\end{algorithm}

\begin{table*}[]
	\centering
	\begin{tabular}{|c|c|c|c|}
		\hline 
		Network type&  Datasets& $\#$ of nodes & $\#$ of edges \\ 
		\hline 
		Social networks& Twitter & $81,306$   & $1,768,149$   \\
		                          & Google+ & $106,674$   & $13,673,453$   \\
		\hline 
		Question answering networks&Askubuntu & $159,316$   & $596,933$   \\
		                          &Mathoverflow & $24,818$   & $239,978$   \\
		\hline
		P2P file sharing networks&p2p-Gnutella & $6,301$ to $62,586$  & $20,777$ to $147,892$ \\
		\hline
		Physics paper citation networks&Cit-HepPh & $34,546$   & $421,578$   \\
		                          &Cit-HepTh& $27,770$   & $352,807$   \\
		\hline
		Friendship networks&Slashdot & $77,360$  & $905,468$ \\
		\hline
		 Wikipedia networks&WikiVote & $7,115$  & $103,689$ \\
		\hline
		 Bitcoin networks&OTC trust weighted signed  & $5,881$  & $35,592$ \\
		\hline
		{Trust networks}&Epinion  & $75,879$  & $508,837$ \\
		\hline
		{Advice-seeking networks} &Advice  & $30$ to $60$  & $200$ to $500$ \\
		\hline
		{Co-sponsorship networks}&US Senate  &$100$ to $104$  &  $4188$ to $6415$\\
		\hline
	\end{tabular} 
	\caption{Summary of static datasets characteristics.} 
	\label{tab:staticdatasets}
\end{table*}

\subsection{Algorithm}
{\emph{gl2vec}} works as follows:  given the topological structure of a directed static or temporal network, we first compute its {graphlet} counts.  For static networks, {we applied JMotif \cite{jmotif} to compute triad counts for networks and random graphs in different null models.  We refer interested readers to \cite{jmotif} for more details.}  For temporal networks, we use the SNAP package \cite{paranjape17} to compute 3-edge, $\delta$-temporal {graphlet} counts.  

We then compute average {graphlet} counts in null models {\bf{\emph{NE}}}. For static networks, there are two approaches: simulation based and probability based.  The simulation based approach generates  a large set of random graphs with the same structure of the given network and a {graphlet} counts are computed for each random graph.  The probability based approach computes the probability of occurrence for each type of graphlet given the in/out degree of the nodes involved.  The time complexity of simulation based approach is $O(MND^2)$, where $M$ is the number of random graphs, $N$ is the number of nodes, and $D$ is the average degree of the graph in consideration.  For probability based approach, the time complexity is $\hat{D}^3$, where $\hat{D}$ is the number of unique bi-degree pair\cite{jmotif}. 

We apply the probability based approach to {\bf{\emph{NE}}} due to its fast computation speed and high accuracy.  We apply the simulation based approach to the \textbf{BDS} null model because it has lower computational complexity.  It also avoids approximation errors that occur when the network size is small (less than $1000$ nodes).  For temporal networks, we generate random graphs by shuffling timestamps on edges and then compute their average temporal {graphlet} counts.  Finally, we compute SRPs for corresponding {graphlets} using Equation~(\ref{eq:srp}). The pseudocode is presented in Algorithm~\ref{alg:bdsAlg}.

\subsection{Experiments}\label{sec:exp}

In this section, we conduct network classification tasks on several real-world static and temporal directed networks. Experiments include two tasks: network type classification and subgraph identification. In network type classification, we use {\emph{gl2vec}} to predict the most {likely purpose of the relation and interaction between nodes, e.g., email communication, question answering or friendship in social networks.}   In subgraph identification,  we predict the community ID for (sub)graphs within the same network. Examples include identifying a department based on email-exchange patterns or detecting a mobile phone user based on their app switching behavior represented as static or temporal networks.

Highlights of our experimental findings include:

$\vartriangleright$ {\emph{gl2vec}}, {constructing vectors for feature representations using static or temporal graphlet SRPs, can outperform or achieve comparable performance with state-of-the-art methods in network type classification.}

 $\vartriangleright$ For static directed networks, the impact of null models on the computation complexity of SRP is significant.  Compared to the other two null models, {\bf\emph{BDS}} and {\bf\emph{MAN}},  {\bf\emph{NE}} has the least control on the network structure, i.e., it provides the most randomness in the structure of a random graph.  Hence, {\bf\emph{NE}} is much easier to implement.  Furthermore, we observe that {\bf\emph{NE}} achieves a slightly better performance in network classification in most cases.

$\vartriangleright$ Adding graphlet features from  {\emph{gl2vec}} to state-of-the-art-methods significantly improves their performance. This suggests that graphlet patterns from SRPs provide substantial additional information about network type that do not exist in state-of-the-art-methods.

$\vartriangleright$ Both static and temporal {graphlets} play important roles in temporal network classification.

\subsubsection{Datasets}\label{sec:datasets}
We use a wide range of real-world network datasets, which only contain topological structure. Attributes of nodes and edges are unknown, except for labels for classification, and timestamps of edges, in temporal networks. These datasets may  challenge some  current state-of-the-art methods that require attributes of nodes or edges.

\noindent\textbf{Static Directed Network Datasets:}
We use different types of static directed networks and perform network classification using their topological structures in our experiments.

\noindent $\bullet$ \textbf{SNAP Datasets \cite{snapnets}:}
For social Networks, \emph{Twitter dataset}  contains $1000$ ego-networks with $81,306$ nodes and $1,768,149$ edges.  \emph{Google+ dataset} contains $133$ ego-networks with $106,674$ nodes and $13,673,453$ edges.  A directed edge from $u$ to $v$ represents that user $u$ follows $v$. The size of ego-networks range from $10$ to $4,964$ nodes. 

\emph{Askubuntu} and \emph{Mathoverflow}  datasets are question-answering networks that store interactions between users.  The interactions include posting answers to question (a2q), comments to questions (c2q) and comments to answers (c2a).  Both datasets contain four directed networks: an a2q network, a c2q network, a c2a network and a network containing all interactions.

\emph{p2p-Gnutella dataset} contains $9$ directed peer-to-peer file sharing networks.  Nodes represent hosts and edges represent topological connections between hosts.

\emph{Cit-HepPh} and \emph{Cit-HepTh} are two physics paper citation networks. \emph{Cit-HepPh} (\emph{Cit-HepTh}). 

\emph{Slashdot} is a friendship network, where users tag each other as friends.  \emph{WikiVote dataset} contains votes from users in Wikipedia to promote other users to become administrators.  \emph{Bitcoin OTC trust weighted signed network} contains ratings from Bitcoin users to other users.

\noindent $\bullet$  \textbf{Other Network Types:}
\emph{Epinion social network} \cite{richardson2003trust} is a who-trusts-whom network from a consumer review site Epinions.com.  A directed edge represents that a user ``trusts'' another user. Advice dataset contains advice-seeking between employees in four different companies \cite{lazega2001collegial,krackhardt1987cognitive,cross2004hidden}.  Co-sponsorship networks \cite{fowler2006connecting} contain US Senate co-sponsorship patterns during the 1995, 2000, 2005, and 2010 congressional terms. Nodes represent senators and a directed edge from $u$ to $v$ represents that senator $u$ cosponsored at least one piece of legislation for which senator $v$ was the primary sponsor.

A summary of characteristics of these datasets are given in Table~\ref{tab:staticdatasets}.

\noindent\textbf{Temporal Directed Network Datasets:}
We also collect temporal directed networks to test feature representation using temporal graphlets.

\noindent $\bullet$ \textbf{Email Networks:} 
EmailEU \cite{Yin17} is a directed temporal network constructed from email exchanges in a large European research institution for a $803$-day period. It contains $986$ email addresses as nodes and $332,334$ emails as edges with timestamps.  There are $42$ ground truth departments in the dataset and we choose $26$ departments whose email network sizes are larger than $10$. EmailTraffic \cite{olsson2011finding} is a temporal directed network storing email interactions of $819$ staff in $23$ different departments in BBN for about $7$ months. Edges with integer timestamps represent emails sent out at a certain time.

We constructed temporal subgraphs, each lasting $12$ weeks for departments in EmailEU networks. This ensures each subgraph becomes a connected network component when converted to an unweighted static graph. We create these graphs at the beginning of every four weeks to avoid too much overlap of edges between graphs. Each department has up to $28$ subgraphs as a result. For departments in EmailTraffic, we create subgraphs at the beginning of every week and each subgraph covers four weeks. 

\noindent $\bullet$  \textbf{SwitchApp:}
(from the Tymer project \cite{turner2019evidence,kun2018aaltd,kun19asonam}) contains application switching data for $53$ Android users over a $42$-day period.  We construct a directed temporal network for each user on each day, where a directed edge (denoted as $e_{uv}$) with an integer {timestamp} $t$ represents a user switching from an app $u$ to another $v$ at time $t$.

\subsubsection{Experiment Setup} 
We compute SRPs for  static and temporal {graphlets} for corresponding static and temporal networks in our datasets. We use three widely used machine learning models that provide good performance using small amounts of training data in multi-class classification: {XGBoosting \cite{chen16}, SVM \cite{cortes1995support}, random forest \cite{svetnik03}.} XGBoosting usually has a superior performance over other classifiers when the dataset is of middle size.  SVM is suitable for a small amount of training data. Random forest not only works well for imbalanced data, but also performs feature selection during training which can help us investigate the usefulness of our feature representation, especially when used in conjunction with other approaches by concatenating the feature vectors.

We use a {grid search method} to search the best hyper-parameters for these models. For the XGBoosting algorithm, the learning rate ranges from $0.001$ to $1$, maximal tree depth range from $4$ to $32$, minimal child weight is $1$ and the subsample ratio of train instances ranges from $0.4$ to $1$. The regularization weight in SVM ranges from $1$ to $8$. In random forest, the number of trees ranges from $50$ to $400$ and the minimal number of samples required to split a tree node from $2$ to $10$. $10$-fold cross-validation is adopted to split the data to select the best parameters.  All experiments are conducted using a cluster with 32 Xeon CPU with 256GB RAM and one Tesla K40 GPU.

We compare the network classification accuracy of {\emph{gl2vec}} to state-of-the-art methods, including graphlet and Weisfeiler-Lehman kernels \cite{shervashidze2011weisfeiler}, and recently developed node and graph embedding methods  \emph{node2vec} \cite{grover2016node2vec}, \emph{struc2vec} \cite{ribeiro17}, \emph{sub2vec} \cite{adhikari2018sub2vec}, \emph{graph2vec} \cite{narayanan2017graph2vec}.   

For node embedding methods such as \emph{node2vec} and \emph{struc2vec}, we apply a sum-based approach \cite{dai2016discriminative} to aggregate node embedding vectors to construct a graph embedding.  We refer interested readers to \cite{hamilton2017representation} for more detail. The length of network embedding (ranging from $50$ to $500$) is determined using grid search and $10$-fold cross-validation.  We modify state-of-the-art methods to apply them to directed graphs: we run a random walk on directed graphs in \emph{sub2vec} instead of undirected graphs.  Some state-of-the-art methods also require node attributes for network embeddings and node degree are suggested for computing undirected graph embeddings \cite{hamilton2017representation}.  For directed networks, we therefore use NetworkX to compute the in/out degree and centralities such as betweenness, closeness and in/out degree centrality for each node.   We also consider additional attributes: counts of subgraphs of a specific triad that a node belongs to. These counts are normalized as a distribution indicating the likelihood a node belongs to a specific triad.

\subsubsection{Network Type Classification Task}
In network type classification, we are given the topological structure of a subgraph in a network.  Our goal is to predict the type of interaction that an edge represents, e.g. email exchange or question answering. 

Among all the datasets introduced in Section~\ref{sec:datasets}, \emph{EmailEU}, \emph{EmailTraffic} and \emph{SwitchApp} datasets have ground truth labels (department ID or user ID) available for each subgraph, which is created from email exchanges in a department or app switch behaviors of a user within a period of time.  Hence, we can obtain all subgraphs for these communities in these three networks.  For the other datasets, there is no ground truth information on network communities; we compute network communities using modularity \cite{newman2006modularity} to obtain subgraphs. These subgraphs are converted into feature vectors using the previously introduced embedding methods and assigned labels according to network types. Finally, we  collect about $10,000$ (sub)graphs from $2,355$ real-world networks taken from $15$ network types introduced above, which include Google+ and Twitter in social networks, high energy physics theory citation networks, Gnutella P2P networks, SwichApp and so on.

\noindent\textbf{Static Directed Networks:}
We use all datasets to evaluate embedding methods on static networks.  Note that we convert temporal networks into unweighted static networks by removing the timestamps on the edges.  Baseline methods include graphlet graph kernel (GK graphlet), Weisfeiler-Lehman graph kernel (GK WL), feature vector with triad distribution (MotifDist), node2vec, graph2vec, sub2Vec and struc2vec.  

We first investigate the impact of null models on our proposed graph representation approach {\emph{gl2vec}}.  This includes the computation of SRPs using the following three different null models: 

(i) \emph{gl2vec(BDS)}: random graphs with the same bidegree sequence;

(ii) \emph{gl2vec(MAN)}: random graphs with the same number of mutual, asymetric and null edges;  

(iii) \emph{gl2vec(NE)}: random graphs with the same number of nodes and edges. 

 The accuracies of {\emph{gl2vec}} with different null models for network type classification are shown in Table~\ref{tab:nullmodelgl2vec}.  We observe that all null models to compute SRP have similar impact on the classification, with {\bf\emph{NE}} performing slightly better than the others. {\bf\emph{NE}} is thus preferable among the three because of its lower computational complexity.  Therefore, in the rest of the paper, we only present results for {\emph{gl2vec}} using {\bf\emph{NE}}, and for simplicity, we denote \emph{gl2vec(NE)} as \emph{gl2vec} directly.  

\begin{table}	
	\centering
	\resizebox{\linewidth}{!}{
	\begin{tabular}{|c|c|c|c|}
		\hline 
		&  XGBoost (\%) & SVM (\%)& RF (\%) \\ 
			\hline 
		gl2vec(BDS)& {\bf 80.92 $\pm$ 3.20} & 72.69 $\pm$3.38 & 80.17 $\pm$4.07 \\ 
		\hline 
		gl2vec(MAN) & {\bf 81.49 $\pm$ 3.33} & 71.39 $\pm$ 3.98 & 79.78 $\pm$3.46  \\  
		\hline 
		gl2vec(NE) & {\bf 81.58 $\pm$3.07} & 71.64 $\pm$2.13 & 79.42 $\pm$3.69 \\
		\hline
	\end{tabular} 
	}
	\caption{The impact of null models on network type classification accuracy of {\emph{gl2vec}}. } 
	\label{tab:nullmodelgl2vec}
\end{table}

\begin{table}	
	\centering
	\resizebox{\linewidth}{!}{
	\begin{tabular}{|c|c|c|c|}
		\hline 
		&  XGBoost (\%) & SVM (\%)& RF (\%) \\ 
		\hline 
		GK Graphlet & {\bf 78.94 $\pm$ 3.18} & 72.66 $\pm$2.79 & 78.72 $\pm$3.01   \\
		 +gl2vec  & 82.18 $\pm$ 2.86& 69.01 $\pm$ 2.27& 81.39 $\pm$ 3.36 \\ 
		\hline 
		GK WL & 78.26 $\pm$2.65  & 72.81  $\pm$ 2.74 & {\bf 78.41 $\pm$3.02}   \\ 
		 +gl2vec  & 82.54 $\pm$ 2.85& 68.59 $\pm$ 2.75& 82.26 $\pm$ 3.43 \\
		\hline 
		MotifDist & {\bf 78.08 $\pm$ 3.34}  & 71.40 $\pm$2.29 &78.01 $\pm$ 3.56   \\
		 +gl2vec & 81.75 $\pm$ 3.48& 69.70 $\pm$ 3.64& 80.95 $\pm$ 3.63 \\ 
		\hline 
		node2vec& {\bf 74.25 $\pm$3.07} & 69.03 $\pm$1.23 & 72.24 $\pm$1.67  \\
		 +gl2vec  & 88.76 $\pm$ 1.26& 73.24 $\pm$ 2.92& 86.14 $\pm$ 1.71 \\ 
		\hline 
		graph2vec& 72.48 $\pm$ 3.99 & 70.81 $\pm$ 3.84 & {\bf 72.61 $\pm$3.36} \\
		 +gl2vec  & 79.83 $\pm$ 4.59& 66.70 $\pm$ 4.04& 80.03 $\pm$ 4.38 \\ 
		\hline 
		sub2vec& {\bf 81.39 $\pm$ 1.70} & 79.69$\pm$ 1.41 & 78.44 $\pm$2.26  \\
		 +gl2vec  & 92.30 $\pm$ 2.29& 83.16 $\pm$ 2.62& 90.01 $\pm$ 2.16 \\ 
		\hline 
		struc2vec& {\bf 79.15 $\pm$ 3.42} & 78.22 $\pm$3.15 & 78.94 $\pm$3.31  \\
		{+nodeTriadDistr} & 81.93 $\pm$ 3.53& 79.18 $\pm$ 3.55& 82.01 $\pm$ 3.42 \\ 
		+gl2vec & 93.38 $\pm$ 1.51& 84.25 $\pm$ 0.82& 93.48 $\pm$ 1.42 \\ 
		\hline 
		gl2vec & {\bf 81.58 $\pm$3.07} & 71.64 $\pm$2.13 & 79.42 $\pm$3.69 \\
		\hline
	\end{tabular} 
	}
	\caption{Network type classification accuracy. We use ``+'' to denote an embedding generated by combining two embedding methods.  Bold indicates the best performaning machine learning model for each embedding. } 
	\label{tab:staticTypeClassification}
\end{table}

The accuracies of different embedding methods for network type classification are presented in Table~\ref{tab:staticTypeClassification}.  We make the following observations:

$\vartriangleright$  The machine learning methods used have an impact on the results.  For this task, XGBoost provides the best performance on average in network type classification. 

$\vartriangleright$ \emph{gl2vec} can outperform or achieve comparable performance with state-of-the-art methods with appropriate machine learning methods.

$\vartriangleright$  Combining \emph{gl2vec} with different state-of-the-art methods by directly concatenating their feature vectors provides significant improvement on the state-of-the-art methods, especially for graph-based network embedding methods sub2vec and struc2vec. This suggests that both our approach and state-of-the-art methods capture important but different features for network type classification, giving the best results through the combination of these features.  Furthermore, this validates the importance of including subgraph information into feature representations for tasks like network classification in which network structure plays a significant role.  
Moreover, there are also improvements on MotifDist and GK Graphlet.  This confirms that adding null models to construct feature representation helps improve prediction performance.  Finally, as representations from {\emph{gl2vec}}, {MotifDist} and {GK Graphlet} all construct features from graphlets, the improvement that {\emph{gl2vec}} brings is not as notable.

\begin{figure}[h]
	\centering
	\includegraphics[width=0.7\linewidth]{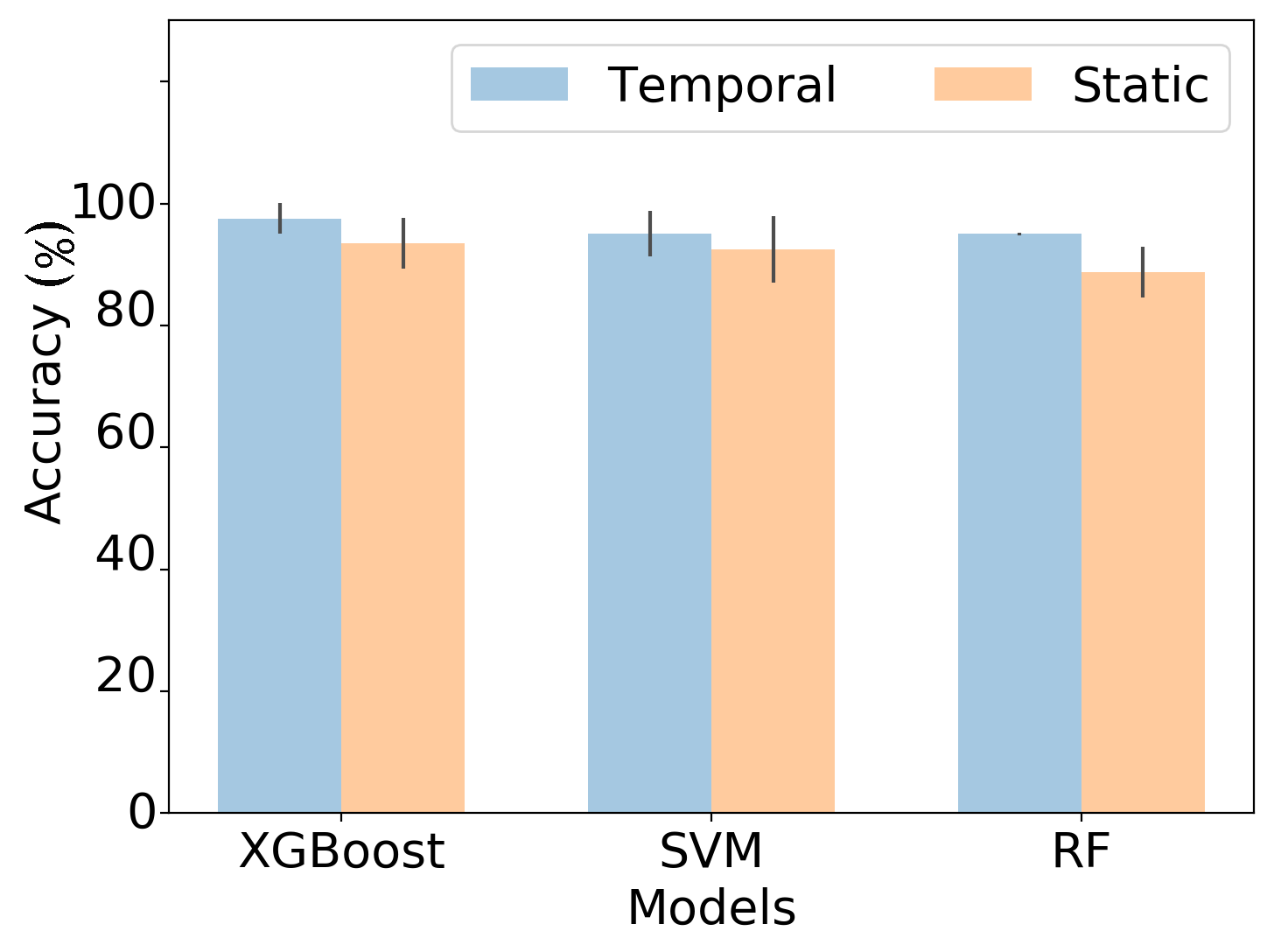}
	\caption{Classifying email datasets and SwitchApp Temporal Networks.}
	\label{fig:networkType}
\end{figure}

\noindent\textbf{Temporal Directed Networks:}
We consider the temporal datasets discussed in Section~\ref{sec:datasets}.  We explore whether temporal {graphlets} provide more information than static {graphlets} in temporal networks. We investigate the effect of graphlets on predicting whether a temporal (sub)graph is an email exchange network or the app switching behavior of a mobile user.   Since the-state-of-the-art methods work only on static networks, we choose \emph{gl2vec(NE)} as a baseline for comparison due to its good performance in static network type classification.   
The results are shown in Figure~\ref{fig:networkType}.  From Figure~\ref{fig:networkType}, we observe that temporal information improves network type classification in all models considered here.  Therefore, it is important to use temporal {graphlets} when constructing vectors for feature representations of temporal networks, since temporal {graphlets} provide more network structure information than static {graphlets}.

\subsubsection{Subgraph Identification Task}
In subgraph identification, we are interested in classifying subgraphs within the same network given their topological structure. For example, we can identify which department an email exchange subgraph belongs to or detect a mobile phone user given their app switching behavior.

We use EmailEU, EmailTraffic and SwitchApp datasets since ground truth labels (department ID or user ID) are available for each subgraph.  We first solve this problem using static graph embedding methods.  Then we investigate whether the timestamp information of edges can help improve identification accuracy.

\noindent\textbf{Static Directed Networks:}
The results on the accuracy of identifications of departments in emailEu, emailTraffic networks and user ID in app switch network using different methods are illustrated in Tables~\ref{tab:EmailEUStatic}, ~\ref{tab:EmailTrafficStatic} and~\ref{tab:SwitchAppStatic}, respectively.  Again, we only present results for {\emph{gl2vec}} using {\bf\emph{NE}} as mentioned earlier.  We cannot obtain results from graph2vec due to its insufficient GPU memory in our test system.   We also evaluate {\emph{gl2vec}} when combined with state-of-the-art methods.  We use ``+" to denote these combinations.  For example, MotifDistr+gl2vec combines feature vectors from MotifDistr and \emph{gl2vec} for feature representation.  

We observe that the addition of graphlet SRP features to state-of-the-art methods can significantly improve performance of the corresponding state-of-the-art methods. This indicates that our {\emph{gl2vec}} method provides new information not present in state-of-the-art methods.

\begin{table}	
	\centering
	\resizebox{\linewidth}{!}{
	\begin{tabular}{|c|c|c|c|}
		\hline 
		 &  XGBoost (\%)& SVM (\%)& RF (\%) \\ 
		\hline {\bf MotifDistr} & 56.68 $\pm$ 6.70 & 45.82 $\pm$ 7.38 & {\bf 61.54 $\pm$ 10.50} \\
			 +gl2vec & 64.18 $\pm$ 6.52& 52.20 $\pm$ 4.80& 63.79 $\pm$ 8.94 \\
		\hline	{\bf GK WL} & 50.96 $\pm$ 8.91 & 47.92 $\pm$ 6.15 & {\bf 57.01 $\pm$ 6.91} \\
			      +gl2vec & 63.12 $\pm$ 5.44& 51.95 $\pm$ 4.44& 65.29 $\pm$ 8.81 \\
		\hline	{\bf GK Graphlet}  & 61.22 $\pm$ 4.70 & 52.32 $\pm$ 5.16 & {\bf 62.90 $\pm$ 4.49} \\
		       +gl2vec & 62.04 $\pm$ 5.69& 52.22 $\pm$ 4.85& 64.35 $\pm$ 8.86 \\
		\hline	{\bf node2vec}  & 52.08 $\pm$ 3.00 & 57.76 $\pm$ 3.11 & {\bf 57.89 $\pm$ 2.83} \\
		       +gl2vec & 63.20 $\pm$ 3.69& 59.20 $\pm$ 5.89& 63.22 $\pm$ 3.40 \\
		\hline	{\bf sub2vec } & 55.45 $\pm$ 3.42 & 52.02 $\pm$ 3.29 & {\bf 59.87 $\pm$ 3.77} \\
		       +gl2vec & 73.01 $\pm$ 8.93& 58.88 $\pm$ 9.42& 77.69 $\pm$ 6.90 \\
		\hline {\bf struc2vec} & 60.25  $\pm$  9.40& 56.8 $\pm$  11.34& {\bf 60.59  $\pm$ 11.14}\\
	       		{+nodeTriadDistr} & 60.78 $\pm$ 9.13& 59.86 $\pm$ 9.22& 61.24 $\pm$ 9.88 \\
		       		+gl2vec & 69.78 $\pm$ 6.20& 54.91 $\pm$ 9.14& 70.30 $\pm$ 7.36 \\
		\hline	
				{\bf gl2vec}& 61.72 $\pm$ 3.09 & 51.03 $\pm$ 3.30 & {\bf 63.09 $\pm$ 3.23}  \\ 

		\hline 
	\end{tabular} }
	\caption{Accuracy in correctly identifying $26$ EmailEU department in static directed networks.}
	\label{tab:EmailEUStatic}
\end{table}

\begin{table}	
	\centering
	\resizebox{\linewidth}{!}{
	\begin{tabular}{|c|c|c|c|}
		\hline 
		&  XGBoost (\%)& SVM(\%) & RF (\%) \\  
		\hline	{\bf MotifDistr} & 67.81 $\pm$ 7.60 & 62.60 $\pm$ 8.27 & {\bf 70.04 $\pm$ 7.48} \\
		       +gl2vec & 78.81 $\pm$ 10.87& 71.93 $\pm$ 6.78& 80.19 $\pm$ 7.73 \\
		\hline	{\bf GK WL}  & 72.18 $\pm$ 5.86 & 70.73 $\pm$ 6.81 & {\bf 75.99 $\pm$ 5.82} \\
		       +gl2vec & 77.96 $\pm$ 9.03& 71.73 $\pm$ 7.45& 80.58 $\pm$ 7.24 \\
		\hline	{\bf GK}  graphlet  & 74.39 $\pm$ 10.71 & 70.77 $\pm$ 12.35 & {\bf 78.61 $\pm$ 8.91} \\
		       +gl2vec & 77.17 $\pm$ 11.73& 71.52 $\pm$ 6.71& 80.18 $\pm$ 7.23 \\
		\hline	{\bf node2vec}  & 74.02 $\pm$ 8.13 & 70.45 $\pm$ 12.25 & {\bf 77.41 $\pm$ 6.93} \\
		       +gl2vec & 85.21 $\pm$ 6.64& 75.36 $\pm$ 6.88& 87.81 $\pm$ 4.93 \\
		\hline	{\bf sub2vec}  & {\bf 77.79 $\pm$ 3.93} & 77.80 $\pm$ 3.63 & 77.01 $\pm$ 3.83 \\
		       +gl2vec & 83.39 $\pm$ 5.43& 86.74 $\pm$ 5.25& 87.00 $\pm$ 4.58 \\
		\hline {\bf struc2vec} & {\bf 73.78  $\pm$  9.40}& 65.33 $\pm$  9.34& 72.16  $\pm$ 9.14\\
				+nodeTriadDistr & 74.35 $\pm$ 10.72& 66.84 $\pm$ 10.08& 77.17 $\pm$ 10.44 \\
				+gl2vec & 79.85 $\pm$ 14.01& 56.23 $\pm$ 14.88& 81.43 $\pm$ 12.38 \\
		\hline
				{\bf gl2vec}  & 76.80 $\pm$ 6.24 & 71.13 $\pm$ 6.49 & {\bf 80.78 $\pm$ 5.65} \\
		\hline 
	\end{tabular} }
		\caption{Accuracy in correctly identifying EmailTraffic department in static directed networks. }
	\label{tab:EmailTrafficStatic}
\end{table}

\begin{table}	
	\centering
	\resizebox{\linewidth}{!}{
	\begin{tabular}{|c|c|c|c|}
		\hline 
		&  XGBoost (\%)& SVM (\%)& RF (\%) \\ 
		\hline	{\bf MotifDistr} & 11.82 $\pm$ 2.03 & 11.62 $\pm$ 2.02 & {\bf 12.33 $\pm$ 2.28} \\
		       +gl2vec & 16.16 $\pm$ 1.85& 12.95 $\pm$ 2.31& 15.34 $\pm$ 1.45 \\
		\hline	{\bf GK WL}  & 11.50 $\pm$ 1.65 & {\bf14.59 $\pm$ 0.97} & 13.43 $\pm$ 2.26 \\
		       +gl2vec & 16.01 $\pm$ 2.43& 13.15 $\pm$ 1.44& 17.31 $\pm$ 1.81 \\
		\hline	{\bf GK graphlet}  & 13.89 $\pm$ 1.26 & 15.24 $\pm$ 1.67 & {\bf 15.29 $\pm$ 2.28} \\
		       +gl2vec & 16.52 $\pm$ 1.77& 13.98 $\pm$ 2.51& 15.95 $\pm$ 1.61 \\
		\hline	{\bf node2vec}  & {\bf 10.15 $\pm$ 1.50} & 7.91 $\pm$ 1.32 & 9.98 $\pm$ 1.66 \\
		       +gl2vec & 16.33 $\pm$ 2.04& 12.94 $\pm$ 2.71& 16.21 $\pm$ 1.97 \\
		\hline	{\bf sub2vec}  & 16.27 $\pm$ 2.20 & 16.19 $\pm$ 4.37 & {\bf 16.54 $\pm$ 1.64} \\
		        +gl2vec & 31.74 $\pm$ 3.58& 23.43 $\pm$ 2.33& 33.94 $\pm$ 4.58 \\
		\hline \textbf{struc2vec} & \textbf{14.18  $\pm$  2.21} & 9.75 $\pm$  2.49& 12.17  $\pm$ 2.64\\
				{+nodeTriadDistr} & 15.23 $\pm$ 2.57& 7.81 $\pm$ 2.02& 13.16 $\pm$ 2.09 \\
				+gl2vec & 19.53 $\pm$ 3.13& 9.30 $\pm$ 1.60& 20.70 $\pm$ 3.07 \\
		\hline
				{\bf gl2vec}  & 16.17 $\pm$ 1.80 & 13.56 $\pm$ 1.60 & {\bf 16.82 $\pm$ 1.31} \\
		\hline 
	\end{tabular} }
	\caption{Accuracy in correctly identifying $53$ SwitchApp user in static directed networks.}
	\label{tab:SwitchAppStatic}
\end{table}

\noindent\textbf{Algorithm Performance with Graphlet Features:}
Tables \ref{tab:EmailEUStatic}, \ref{tab:EmailTrafficStatic} and \ref{tab:SwitchAppStatic} indicate that random forest (RF) is usually more accurate for graph embeddings that include our SRP feature vectors in baselines. This is because RF automatically performs feature selection during training and adapts to the change in number of features. As a result, it is easier for RF to achieve better results given a similar amount of effort fine-tuning the hyper-parameters. Finally, the improvements on all machine learning models confirm that it is worth combining our {\emph{gl2vec}} graph embedding with other methods to achieve better performance.

\noindent\textbf{Temporal Directed Networks:}
 In temporal networks from EmailEU and EmailTraffic, we attempt to identify which department emails belong to. For the SwitchApp dataset, we attempt to identify a particular user based on their daily app switching behavior represented as a temporal network.

For the EmailEU and EmailTraffic dataset, multiple temporal and static networks are constructed for each department from email exchanges as described in Section~\ref{sec:datasets}.  For the SwitchApp dataset, $42$ temporal and static networks are generated for each person from their app switching behaviors every day.  XGBoosting, SVM and random forest are implemented using different network feature representations: subgraph ratio profile (SRP) with temporal (``Temporal'') and with static (``Static") graphlets, combined SRPs with both temporal and static graphlet (``Temp+Static"). We illustrate the result from sub2vec representation (``Sub2Vec") because it performs best among the baseline methods.  Finally, we create a combination of all three representations  (``CombineAll").

\begin{figure}[h]
	\centering
	 \includegraphics[width=0.7\linewidth]{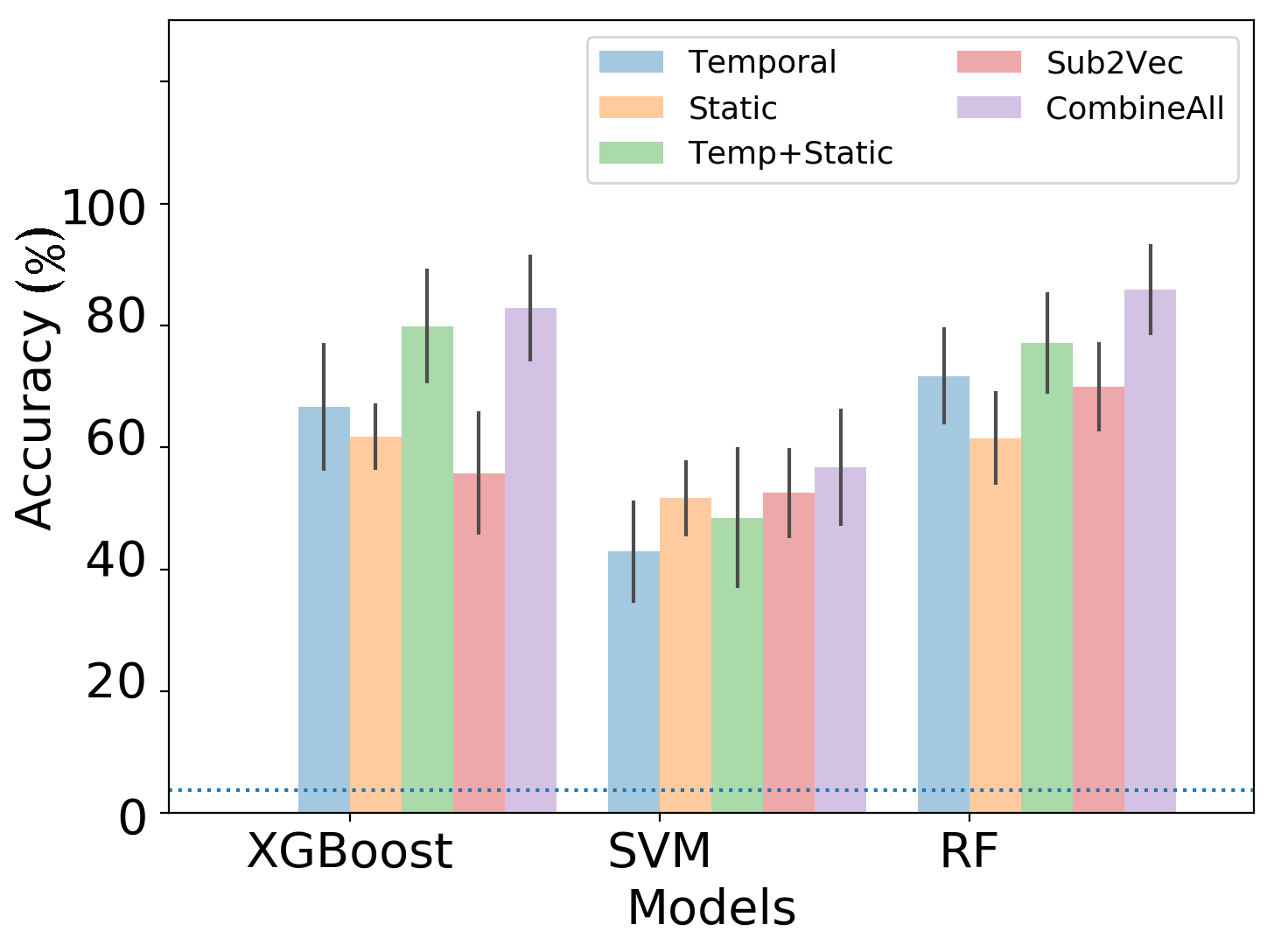}
 \caption{Department identification in EmailEU dataset. Dashed line represents the accuracy of a random selection model.}
  \label{fig:departmentIdentification}
\end{figure}

\begin{figure}[h]
	\centering
	  \includegraphics[width=0.7\linewidth]{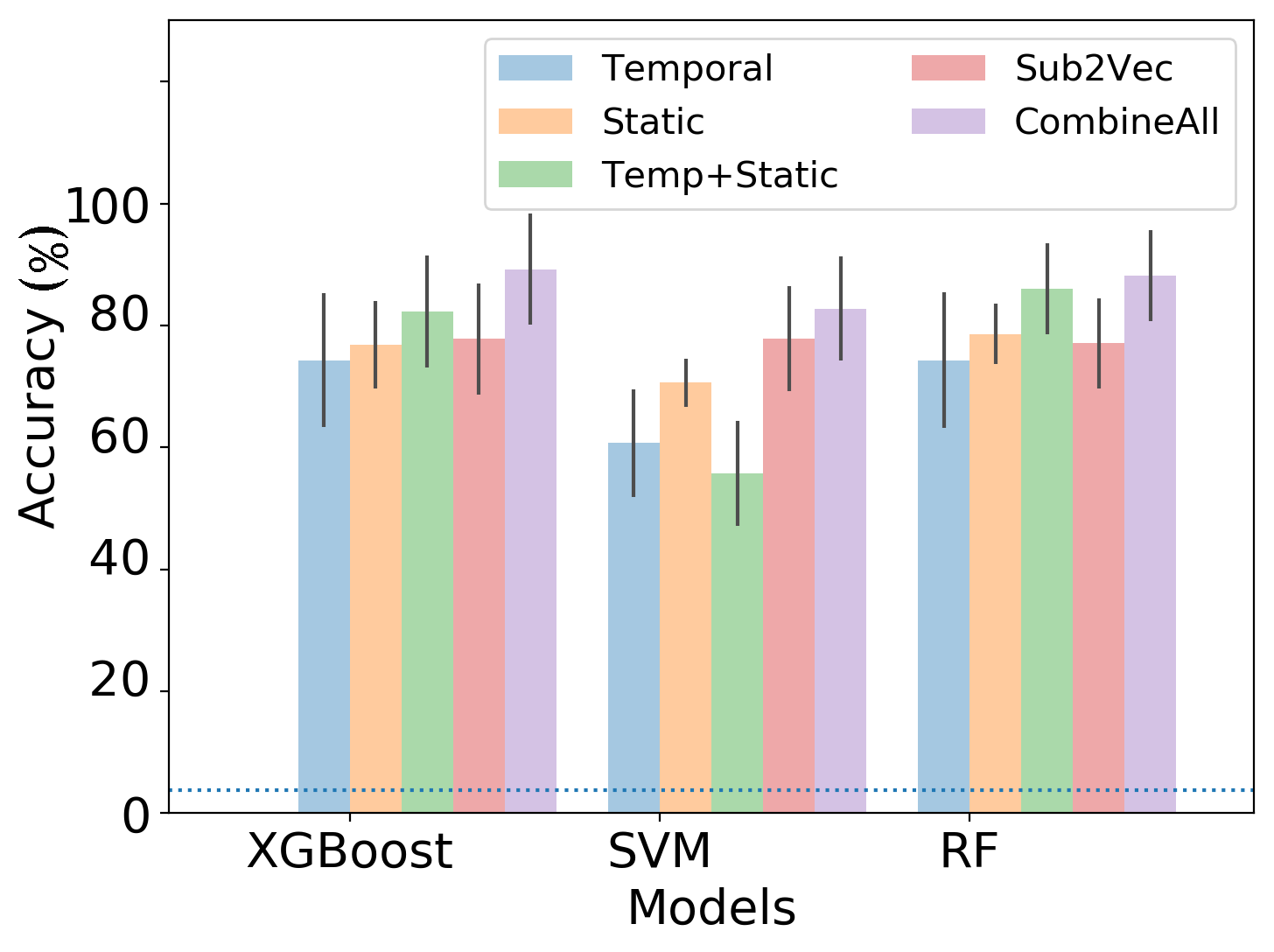}
  \caption{BBN department identification in EmailTraffic. Dashed line represents the accuracy of a random selection model.}
  \label{fig:trafficIdentification}
\end{figure}

\begin{figure}[h]
	\centering
	 \includegraphics[width=0.7\linewidth]{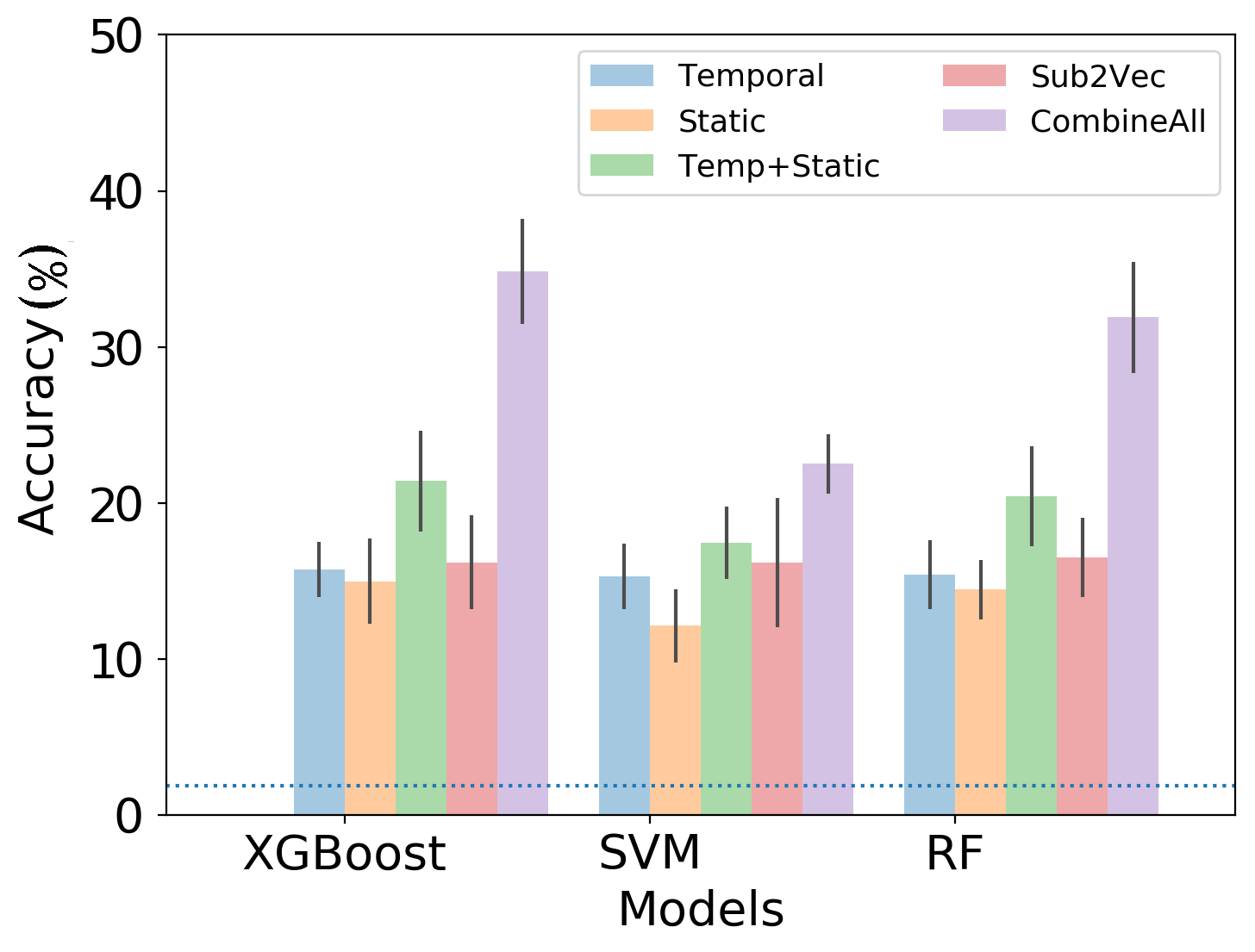}
	\caption{User identification in SwitchApp. Dashed line represents the accuracy of a random selection model. }
	\label{fig:userIdentification}
\end{figure}

The results for EmailEU, EmailTraffic and SwitchApp are shown in Figures~\ref{fig:departmentIdentification}, ~\ref{fig:trafficIdentification}, and~\ref{fig:userIdentification}, respectively. The dashed line is the accuracy of a random selection model. We make the following observations:

$\vartriangleright$ The accuracy achieved by temporal {graphlet} embedding is slightly better than that of static {graphlet} embedding in both emailEU and SwitchApp datasets.  However, static {graphlet} embedding performs better than temporal {graphlets} in EmailTraffic dataset. This shows that static {graphlets} are still useful for temporal network classification and can capture useful features even better than temporal {graphlets} in some datasets. Hence, we combine both static and temporal {graphlet} features (``Temp+Static") and observe that this achieves a significant improvement in accuracy against the baseline,  which suggests that both temporal and static graphlets are useful for network identification (of departments or personal app switching behavior).   

$\vartriangleright$ Our {graphlet}-based network embeddings are competitive with the state-of-the-art method, \emph{sub2vec}.  

$\vartriangleright$ Combining all three graph embedding vectors (static and temporal graphlets, and \emph{sub2vec}) for classification yields the best accuracy.  This suggests that both our static and temporal embedding approaches capture useful features to boost the performances of state-of-the-art methods.  

$\vartriangleright$ It should be noted that the accuracy for SwitchApp is much lower than the other two datasets.  This is primarily because all users have very structurally similar app switch networks \cite{turner2019evidence} and therefore the ability to distinguish between them is a challenging task, combined with the large number of classification choices.

\section{gl-DCNN: Diffusion-Convolutional Neural Network with Graphlets}\label{sec:gcnn}

Diffusion-convolutional neural networks (DCNNs) \cite{atwood2016diffusion} have been proposed as a new model for graph-structured data, which significantly outperforms state-of-the-art methods for node classification tasks, and offer comparable performance to baseline methods for graph classification tasks.   In this section, we argue that the subgraph features can also significantly improve the performance of DCNN in network classification tasks, especially for graph classification.

\subsection{Node Features from Graphlets}
DCNNs require node attributes to embed nodes and graphs. We generate additional node attributes that contain graphlet information.   Let $v$ be a node in a graph of size $n$, we have that $v$ belongs to $\frac{(n-1)(n-2)}{2}$ unique triad ($3$-node subgraphs).   We compute the distribution of all $16$ triad types (see Figure~\ref{fig:static-motifs}) in those subgraphs as attributes for node $v$.

\begin{table}	
	\centering
	\resizebox{\linewidth}{!}{
	\begin{tabular}{|c|c|c|c|}
	\hline 
	 \multicolumn{4}{|c|}{Cora} \\
		\hline 
		Model &  Accuracy (\%) & micro $F_1$ (\%)& macro $F_1$ (\%) \\ 
		\hline 
		DCNN& 86.77  & 86.77 & 85.84   \\
		+gl & 87.21 & 87.21 & 86.07 \\
		\hline 
	\end{tabular} 
	}
	\label{tab:cora}
\end{table}
\begin{table}	
	\centering
	\resizebox{\linewidth}{!}{
	\begin{tabular}{|c|c|c|c|}
	\hline 
	 \multicolumn{4}{|c|}{Pubmed} \\
		\hline 
		Model &  Accuracy (\%) & micro $F_1$ (\%)& macro $F_1$ (\%) \\ 
		\hline 
		DCNN& 89.73  & 89.73 & 89.42   \\
		+gl & 90.66 & 90.66 & 89.82 \\
		\hline 
	\end{tabular} 
	}
	\caption{Node classification for Cora paper citation network and Pubmed network.  We use ``+'' to denote an embedding generated by combining features generated from graphlets. When subgraph feature is added (+gl), the performance is improved. } 
	\label{tab:pubmed}
\end{table}

\subsection{Node Classification}
We report classification accuracy as well as micro- and macro-averaged $F_1,$ where $F_1$ score is a weighted average of precision and recall, computed by 
$$F_1= \frac{2\times\text{precision}\times \text{recall}}{\text{precision} +\text{recall}}. $$
Micro $F_1$ calculates $F_1$ score globally by counting the total true positives, false negatives and false positives, while macro $F_1$ calculates $F_1$ score for each label, and find their unweighted mean. This does not take label imbalance into account.   We report each measure as a mean and confidence interval computed from several trials.

\subsubsection{Node Classification Data}
The Cora paper citation network \cite{sen2008collective} contains $2,708$  nodes representing machine learning papers  and $5,429$ edges representing citations. Each paper has a label from seven machine learning subjects.  The Pubmed network \cite{sen2008collective} contains $19,717$ nodes representing scientific papers on diabetes and $44,338$ links.  Each paper is assigned with a label from three classes, where the label is the type of diabetes addressed in the publication. 

DCNN \cite{atwood2016diffusion} considers both Cora and Pubmed networks as undirected graphs, however, we treat them as directed graphs in our embedding method.  {Each undirected edge is considered as two reciprocal edges.}  To characterize the importance of subgraph features (graphlets), we concatenate the feature vectors for nodes generated from graphlets to the counterparts of DCNN for classification.

\subsubsection{Result Discussion}
Table~\ref{tab:pubmed} compares the performance of a pure DCNN and a DCNN concatenated with subgraph features generated from graphlets for node classification.   It is clear that subgraph features improves the performance in classification accuracy, and micro- and macro-averaged $F_1$ scores.  This indicates that our subgraph-based node embedding is representing more information than is captured by a DCNN directly.  However, we note that the improvement is not that significant, one reason for this is that the count for most subgraphs are close to zero in the datasets we used here, and the subgraph distribution is similar in the citation network.

\subsection{Graph Classification}
We also ran experiments to investigate how subgraph features improve DCNN performance when labels are learned on the whole network.  

\subsubsection{Graph Classification Data}
Similar to Atwood and Towsley \cite{atwood2016diffusion}, we consider a standard set of graph classification datasets including NCI1, NCI109, MUTAG, PCI and ENZYMES.  There are $4,100$ and $4,127$ graphs in NCI1 and NCI109 \cite{wale2008comparison} that represent chemical compounds.  The label of each graph is decided by whether it has the ability to suppress or inhibit the growth of a panel of human tumor cell lines, and the label of each node is chosen from $37$ possible labels for NCI1, and $38$ possible labels for NCI109.  There are $188$ nitro compounds in MUTAG \cite{debnath1991structure} which are labeled as either aromatic or heteroaromatic with seven node features.  The $344$ compounds in PTC \cite{toivonen2003statistical} are labeled with whether they are carcinogenic with rats with $19$ nodes features.  Three node features are associated with $600$ proteins in ENZYMES \cite{borgwardt2005protein}.

\begin{table}	
	\centering
	\begin{tabular}{|c|c|c|}
		\hline 
		Dataset & Model  & Accuracy (\%)  \\ 
		\hline 
		NCI1 &DCNN&  62.61 $\pm$ 3.23   \\
		&+gl &   65.83 $\pm$ 4.41\\
		\hline 
		NCl109 &DCNN&  61.72 $\pm$ 3.10   \\
		&+gl &   62.13 $\pm$ 3.43\\
		\hline 
		MUTAG &DCNN&  67.44 $\pm$ 2.84   \\
		&+gl &   69.49 $\pm$ 2.75\\
		\hline 
		PCI &DCNN&  55.46 $\pm$ 3.43   \\
		&+gl &   55.83 $\pm$ 3.11\\
		\hline 
		ENZYMES &DCNN&  18.16 $\pm$ 2.49   \\
		&+gl &   18.01 $\pm$ 1.89\\
		\hline 
	\end{tabular} 
	\caption{Graph classification comparison of performance between a pure DCNN and a DCNN concatenated with subgraph features from  graphlets.   We use ``+'' to denote an embedding generated by combining features generated from graphlets. } 
	\label{tab:graphclassification-dcnn}
\end{table}

\subsubsection{Result Discussion}
Table~\ref{tab:graphclassification-dcnn} compares the performance of pure a DCNN \cite{atwood2016diffusion} and a DCNN concatenated with subgraph features generated from graphlets.   In contrast with node classification experiments, we observe that subgraph features can significantly improve the performance for graph classification, especially for NCI1, NCI109 and MUTAG datasets.   These results further suggest that graphlet features play a critical role in graph classification and are helpful for improving classification accuracy.

\section{Related Work}\label{sec:related}

The primary focus of related works in classifying networks involves examining the topological structure of the graph. The work most related to our method is graph kernel, which has been used to calculate similarities between static undirected graphs \cite{gauzere2015treelet,kaspar2010graph,yanardag15}.  However, the corresponding computational complexity grows significantly with increase in network size.  Moreover, studies in graphlet kernel do not consider features generated by comparing graphlet counts between an empirical network and random graphs from different null models.  In our experiments we have established that different null models can yield significant improvement in network classification. 
 
Different node embedding techniques have been proposed in recent years, such as node2Vec \cite{grover16}, DeepWalk \cite{perozzi14}, Line \cite{tang15} and Local Linear Embedding \cite{roweis00} that use feature vectors to embed nodes into high-dimensional space and empirically perform well.  However, these methods can only be applied to node classification but not graph classification.   Neural networks on graphs, such as Graph neural network (GCN) \cite{kipf2016semi, defferrard2016convolutional}, also obtain the embedding for each node.  It has been recently shown that GCN-based approaches obtain competitive results against kernel-based methods and graph-based regularization techniques, e.g., \cite{tu2018deep} exploits structural regularity regarding the regular equivalence, \cite{pan2018adversarially} considers embedding distribution, or heterogeneous information is incorporated in \cite{huang2017label,gao2018deep}.  However, these methods are usually computationally expensive and used for small scale tasks.

Additionally, several approaches have been proposed to aggregate node feature vectors to a feature vector for networks.  For example, a graph-coarsening approach \cite{defferrard2016convolutional} computes a hierarchical structure containing multiple layers; nodes in lower layers are clustered and combined as a node in upper layers using element-wise max-pooling. However, this has high computational complexity.  
Some approaches \cite{niepert2016learning} define an order of nodes and concatenate their feature vectors for a convolutional neural network for classification, however, this can only be applied to undirected static networks.  {Recently, some subgraph embedding based approaches were proposed.  \emph{struc2vec} \cite{ribeiro17} applied sum-based approach such as mean-field \cite{dai2016discriminative}  and loopy belief propagation \cite{murphy1999loopy} to aggregate node embedding to graph representation. \emph{sub2vec} \cite{adhikari2018sub2vec} embedded subgraphs with arbitrary structure, while \emph{graph2vec} \cite{narayanan2017graph2vec} was proposed based on a doc2vec framework to learn data-driven distributed representations of arbitrary sized graphs. But these embeddings do not fully capture network structures to the best performance.  Similar to ours is \cite{allan2009using}, which uses motif frequencies, while we use graphlet distributions and SRP.  Furthermore, \cite{allan2009using} requires node labels, but we do not with {\emph{gl2vec}} and {\emph{gl-DCNN}}. 

\section{Conclusion}\label{sec:conclusion}

We presented two novel graphlet-based technologies, \emph{gl2vec} for network embedding and \emph{gl-DCNN} for diffusion-convolutional neural networks.  We have shown that the features learned from graphlets can significantly improve the performance of state-of-the-art methods in network analysis.  Firstly, examined the utility of {\emph{gl2vec}} by using its features in different network classification tasks for both static and temporal directed networks and found that concatenating \emph{gl2vec} with many state-of-the-art methods yields the best accuracy for real-world applications such as identifying network types, predicting community ID for subgraphs and detecting mobile phone users based on their app-switching behaviors.  Secondly, we showed that additional node attributes learned from graphlet features can significantly improve the performance of DCNN, especially in graph classification.   In future work, we will investigate if graphlet census information can serve as features for nodes in a network.  Specifically, we will investigate whether embedding each node with the numbers of graphlets that it belongs to in a network can improve node and network classification further.

\bibliographystyle{abbrv}
\bibliography{refs}

\begin{IEEEbiography}[{\includegraphics[width=1.05in,height=1.45in,keepaspectratio]{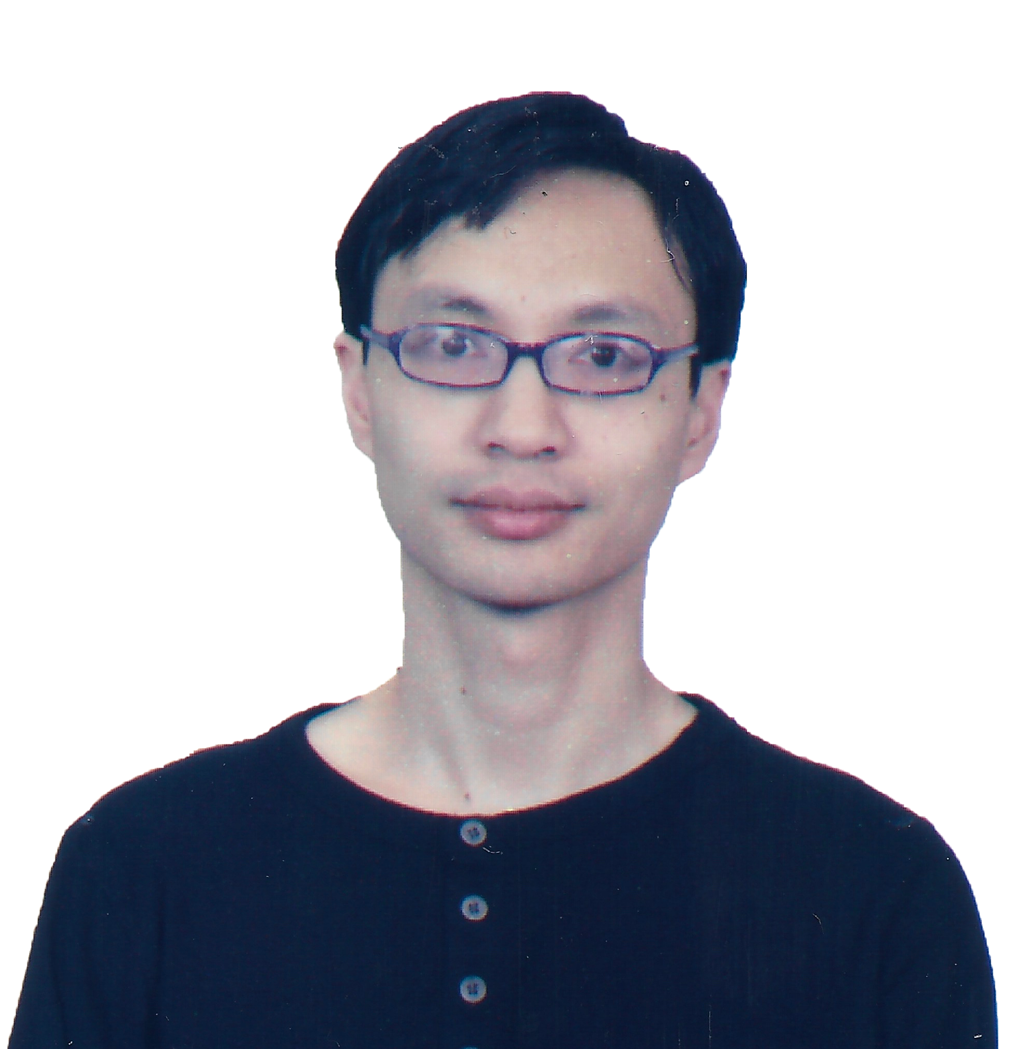}}]{Kun Tu} holds a Ph.D. (2019) in Computer Science from the University of Massachusetts Amherst. He is currently working as a Machine Learning Scientist, designing and developing machine learning models based on large scale industry problems and deploying those models in parallel. His researching interest focuses on Network Science, Machine Learning and Big Data.

\end{IEEEbiography}

\begin{IEEEbiography}[{\includegraphics[width=1.05in,height=1.45in,keepaspectratio]{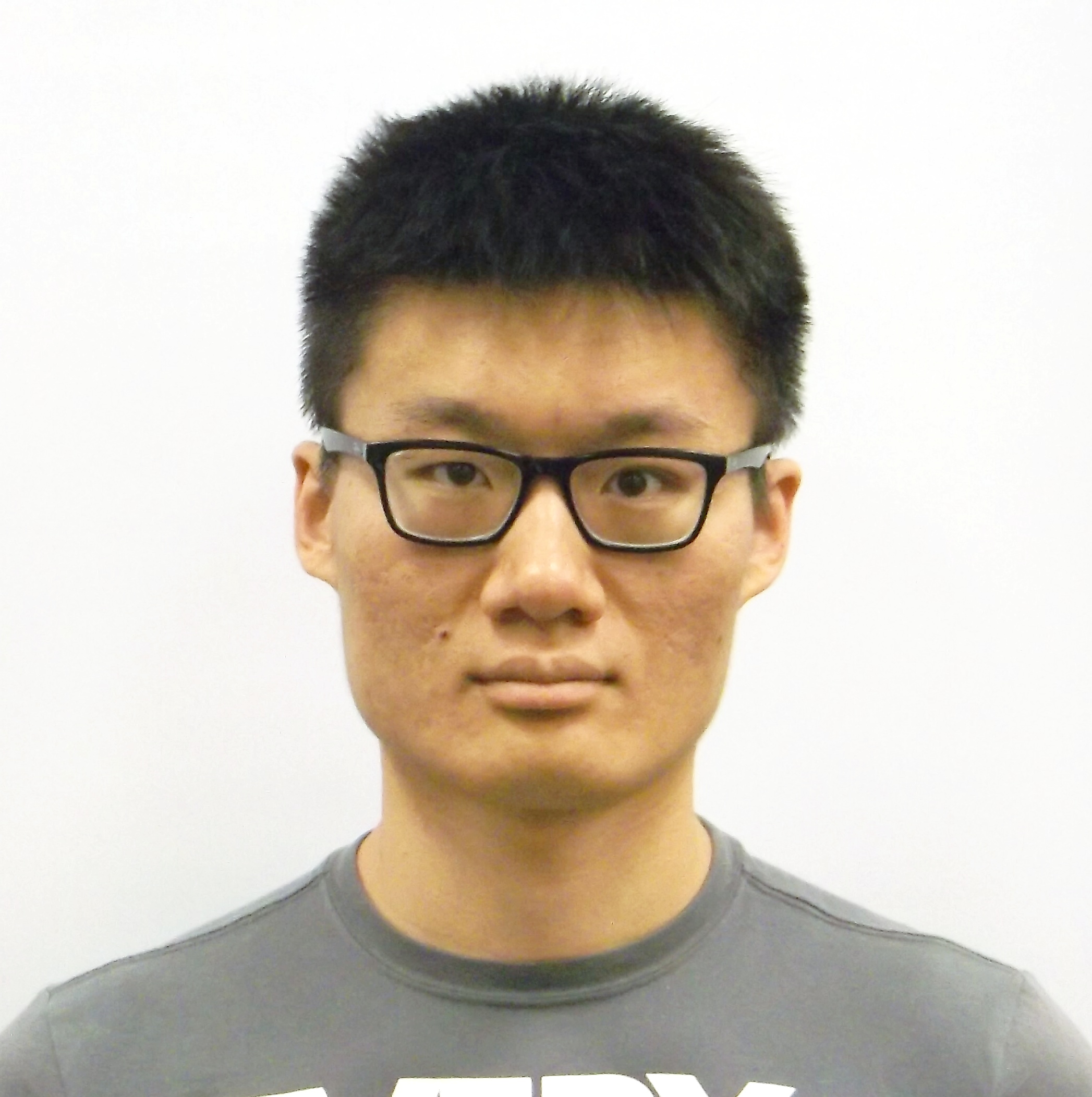}}]{Jian Li}
 is an Assistant Professor of Computer Engineering in Department of Electrical and Computer Engineering at Binghamton University, the State University of New York.  He was a postdoctoral research associate at the University of Massachusetts Amherst.  He received his Ph.D. in Computer Engineering from Texas A\&M University in December, 2016, and B.E. in Electrical Engineering from Shanghai Jiao Tong University in June, 2012.  His current research interests lie broadly in the interplay of large scale networked systems and big data analytics focusing on applying machine learning, online algorithms, game theory, and signal processing technologies to big data applications. 
\end{IEEEbiography}

\begin{IEEEbiography}[{\includegraphics[width=1.05in,height=1.45in,keepaspectratio]{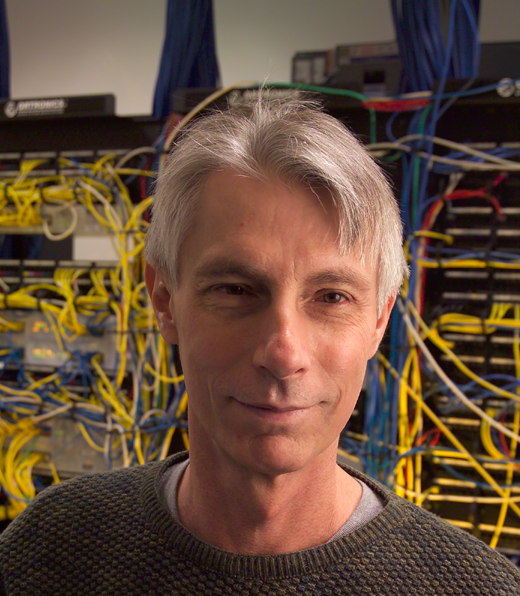}}]{Don Towsley}
holds a B.A. in Physics (1971) and a Ph.D. in Computer Science (1975) from University of Texas.  He is currently a Distinguished Professor at the University of Massachusetts in the College of Information \& Computer Sciences.  He has held visiting positions at numerous universities and research labs. His research interests include network science, performance evaluation, and quantum networking.

He is co-founder and Co-EiC of the new ACM Transactions on Modeling and Performance Evaluation of Computing Systems (TOMPECS), and has served as Editor-in-Chief of {\em IEEE/ACM Transactions on Networking} and on numerous editorial boards.  He has served as Program Co-chair of several conferences including INFOCOM 2009.

He is a corresponding member of the Brazilian Academy of Sciences and has received several achievement awards including the 2007 IEEE Koji Kobayashi Award and the 2011 INFOCOM Achievement Award. He has received numerous paper awards including the 2012 ACM SIGMETRICS Test-of-Time Award, a 2008 SIGCOMM Test-of-Time Paper Award, and a 2018 SIGMOBILE Test-of-time Award.  He also received the 1998 IEEE Communications Society William Bennett Best Paper Award. Last, he has been elected Fellow of both the ACM and IEEE.	
\end{IEEEbiography}

\begin{IEEEbiography}[{\includegraphics[width=1.05in,height=1.45in,keepaspectratio]{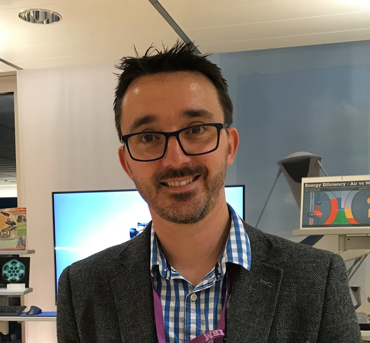}}]{Dave Braines} 
is the Chief Technology Officer for Emerging Technology, IBM Research UK, and is a Fellow of the British Computer Society.  As a member of IBM Research he is an active researcher in the field of Artificial Intelligence and is currently focused on Machine Learning, Deep Learning and Network Motif analysis.  He has published over 100 conference and journal papers and is currently the industry technical leader for a 10-year research consortium comprised of 17 academic, industry and government organisations from the UK and US.  Dave is passionate about human-machine cognitive interfaces and has developed a number of techniques to support deep interactions between human users and machine agents.

Since 2017 Dave has been pursuing a part-time PhD in Artificial Intelligence at Cardiff University, and in his spare time he likes to get outdoors for camping, walking, kayaking, cycling or anything else that gets him away from desks and screens!

\end{IEEEbiography}

\begin{IEEEbiography}[{\includegraphics[width=1.05in,height=1.45in,keepaspectratio]{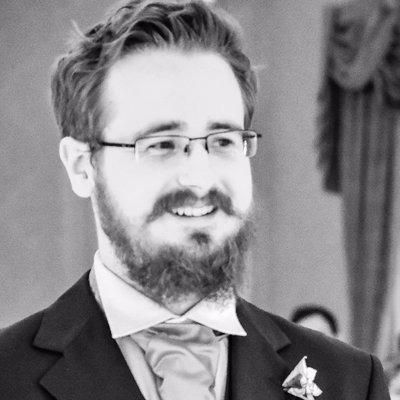}}]{Liam D. Turner}
is an Assistant Professor of Computer Science at the School of Computer Science and Informatics, Cardiff Univeristy, UK. He received his Ph.D. from Cardiff University in 2017 and B.Sc in Computer Science from Cardiff University in 2013. His current research interests primarily surround using computational models to represent and understand complex systems and behaviors and how machine learning can be leveraged to develop systems for supporting human behavior.

\end{IEEEbiography}

\end{document}